\documentclass[twocolumn,showpacs,aps,prd]{revtex4}
\usepackage{graphicx}
\usepackage{dcolumn}
\usepackage{amsmath}
\usepackage{epsfig}

\RequirePackage{xspace}

\newcommand{\gev}{\ensuremath{\mathrm{\,Ge\kern -0.1em V}}\xspace}
\newcommand{\mev}{\ensuremath{\mathrm{\,Me\kern -0.1em V}}\xspace}
\newcommand{\mevcc}{\ensuremath{{\mathrm{\,Me\kern -0.1em V\!/}c^2}}\xspace}
\newcommand{\dedx}{\ensuremath{\mathrm{d}\hspace{-0.1em}E/\mathrm{d}x}\xspace}
\newcommand{\jprBase}        {Phys.\ Rev.\xspace}

\newcommand{\jprd}      [1]  {\jprBase\ D~{\bf #1}}

\def\epem       {\ensuremath{e^+e^-}\xspace}

\def\piz        {\ensuremath{\pi^0}\xspace}
\def\pip        {\ensuremath{\pi^+}\xspace}
\def\pim        {\ensuremath{\pi^-}\xspace}
\def\KS         {\ensuremath{K^0_{\scriptscriptstyle S}}\xspace} 
\def\jpsi       {\ensuremath{{J\mskip -3mu/\mskip -2mu\psi\mskip 2mu}}\xspace}
\def\chic#1{\ensuremath{\chi_{c#1}}\xspace} 
\def\invpb      {\ensuremath{\mbox{\,pb}^{-1}}\xspace}
\def\fourpiz    {\ensuremath{\piz\piz\piz\piz}\xspace}
\def\psiprime   {\ensuremath{\psi^\prime}\xspace}
\def\fz#1       {\ensuremath{f_0({#1})}\xspace} 
\def\Erad       {\ensuremath{E^*_\gamma}\xspace}
\def\besiii     {BESIII\xspace}

\def\etal       {{\em et al.}\xspace}

\begin{document}

\title{
{\boldmath First Observation of the Decays $\chic{J} \to\fourpiz$}
}
\date{\today}

\author{
\small
M.~Ablikim$^{1}$, M.~N.~Achasov$^{5}$, L.~An$^{9}$, Q.~An$^{36}$, Z.~H.~An$^{1}$, J.~Z.~Bai$^{1}$, R.~Baldini$^{17}$, Y.~Ban$^{23}$, J.~Becker$^{2}$, N.~Berger$^{1}$
% ...
\\
\vspace{0.2cm}
(BESIII Collaboration)\\
\vspace{0.2cm} {\it
$^{1}$ Institute of High Energy Physics, Beijing 100049, P. R. China\\
$^{2}$ Bochum Ruhr-University, 44780 Bochum, Germany\\
% ...
$^{5}$ G.I. Budker Institute of Nuclear Physics SB RAS (BINP), Novosibirsk 630090, Russia\\
% ...
$^{9}$ Guangxi University, Naning 530004, P. R. China\\
% ...
$^{17}$ Laboratori Nazionali di Frascati - INFN, 00044 Frascati, Italy\\
% ...
$^{23}$ Peking University, Beijing 100871, P. R. China\\
% ...
$^{36}$ University of Science and Technology of China, Hefei 230026, P. R. China\\
% ...
}
}

\newpage

\begin{abstract}
We present a study of the $P$-wave spin -triplet charmonium \chic{J} decays ($J=0,1,2$) into \fourpiz. The
analysis is based on 106 million \psiprime decays recorded with the
\besiii detector at the BEPCII electron positron collider. 
The decay into the \fourpiz hadronic final state is
observed for the first time. We measure the branching fractions
${B}(\chic{0}\to\fourpiz)=(3.34\pm 0.06 \pm 0.44)\times10^{-3}$,
${B}(\chic{1}\to\fourpiz)=(0.57\pm 0.03 \pm 0.08)\times10^{-3}$, and
${B}(\chic{2}\to\fourpiz)=(1.21\pm 0.05 \pm 0.16)\times10^{-3}$, where
the uncertainties are statistical and systematical, respectively.
\end{abstract}


\maketitle

\section{Introduction}
In the quark model, the \chic{J} ($J=0,1,2$) mesons are the $^3P_J$ charmonium
states. Their decays are experimentally and theoretically not as well
studied as the vector charmonium states \jpsi and \psiprime. 
In contrast to the latter ones,
$\chi_{cJ}$ cannot be produced directly in $\epem$
annihilation. However, they can
be produced in radiative decays $\psiprime\to\gamma\chic{J}$, providing
a clean environment to study their decays. 

Recent theoretical work indicates that the Color Octet Mechanism
\cite{com} could have large contributions to the decays of the
$P$-wave charmonium states. However, these calculations as well as
experimental measurements still have large errors and thus more
precise experimental data besides more theoretical efforts are
mandatory to further understand \chic{J} decay dynamics.  Furthermore,
the \chic{0} and \chic{2} states are expected to annihilate via
two-gluon processes into light hadrons and may therefore allow the
study of glueball dynamics. Thus the measurement of as many
exclusive hadronic $\chi_{cJ}$ decays as possible is valuable.

The $\chi_{cJ}$ decays into four pions have the largest branching fractions
among the known hadronic $\chi_{cJ}$ decay modes \cite{pdg}.
Presently only the decays into $\pi^{+}\pi^{-}\pi^{+}\pi^{-}$ 
and into $\pi^{+}\pi^{-}\pi^{0}\pi^{0}$ are measured by previous experiments.  
The branching fractions are shown in Table \ref{tab:bf}.
In this paper, we present a study of exclusive $\chi_{cJ}$ decays into
\fourpiz.

\begin{table}[tb]
\begin{center}
\begin{tabular} {lr}
\hline\hline
channel & branching fraction [\%]\\\hline
$\chic{0}\to \pip\pim\pip\pim$ & $2.27\pm 0.19$ \\
$\chic{1}\to \pip\pim\pip\pim$ & $0.76\pm 0.26$ \\
$\chic{2}\to \pip\pim\pip\pim$ & $1.11\pm 0.11$ \\
\hline
$\chic{0}\to \pip\pim\piz\piz$ & $3.4\pm 0.4 $ \\
$\chic{1}\to \pip\pim\piz\piz$ & $1.26\pm 0.17 $ \\
$\chic{2}\to \pip\pim\piz\piz$ & $2.00\pm 0.26$ \\
\hline \hline
\end{tabular}
\caption{Branching fractions of $\chi_{cJ}$ into $\pi^{+}\pi^{-}\pi^{+}\pi^{-}$ and $\pi^{+}\pi^{-}\pi^{0}\pi^{0}$ \cite{pdg}.}
\label{tab:bf}
\end{center}
\end{table}

\section{The \besiii Experiment and Data Set}
\label{sec:detector}
We use a data sample of about 106
million \psiprime decays recorded with the \besiii
detector~\cite{besnim} at the energy-symmetric double-ring \epem
collider BEPCII~\cite{bepcii}. The primary data sample corresponds to
an integrated luminosity of 156.4\invpb collected at the peak of the
\psiprime resonance. In addition, a 42.6\invpb data sample collected
about 36\mev below the resonance is used for background studies.

The \besiii detector is described in detail elsewhere \cite{besnim}.
Charged particle momenta are measured with a small-celled, helium
gas-based main drift chamber with 43 layers operating within the $1~\mathrm{T}$
magnetic field of a solenoidal superconducting magnet. Charged
particle identification is provided by measurements of the specific
ionization energy loss \dedx in the tracking device and by means of a plastic scintillator Time of Flight system composed of a barrel part and two end caps. 
Photons are detected and their energies and
positions measured with an electromagnetic calorimeter (EMC) consisting of
6240 CsI(Tl) crystals arranged in a barrel and two end caps. The
return yoke of the magnet is instrumented with Resistive Plate
Chambers arranged in 9 (barrel) and 8 layers (end caps) for
discrimination of muons and charged hadrons.

\section{Data Selection} 
\label{sec:selection}
\begin{figure}[tb]
\includegraphics[width=\columnwidth]{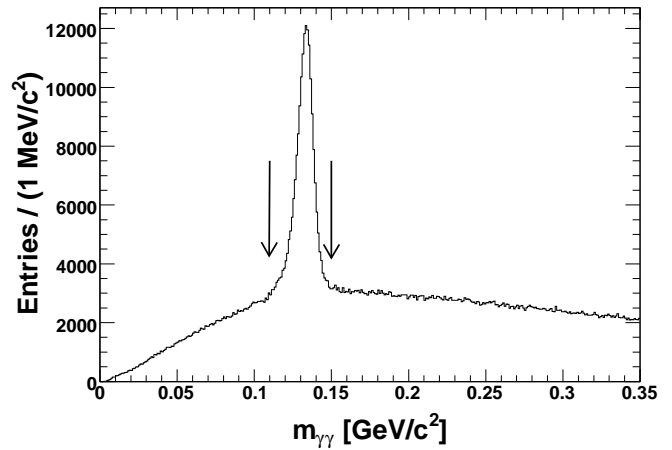}
\caption{Invariant $\gamma\gamma$ mass distribution of all photon pair combinations for selected $9\gamma$ events. The arrows indicate the mass window used for the selection of \piz candidates.}
\label{fig:invgg_pi0}
\end{figure}

We reconstruct the entire event from the decay chain of the charmonium
transitions $\psiprime\to\gamma\chic{J}$ followed by the hadronic
decays $\chic{J}\to\fourpiz$. A photon candidate is defined as a
shower detected with the EMC exceeding an energy deposit of 25\mev in
the barrel region (covering the region $|\cos\theta|<0.8$ of the polar
angle) and of 50\mev in the end caps ($0.86<|\cos\theta|<0.92$). The
average event vertex of each run is assumed as the origin of these
candidates. We restrict the analysis to events having nine photon
candidates and no reconstructed charged particle.  The energy sum of
the nine photons must be within the range 3.45-3.80\gev.  To
reconstruct $\piz\to\gamma\gamma$ candidates we use pairs of photon
candidates, having an invariant mass between 110 and 150\mevcc.
Fig.~\ref{fig:invgg_pi0} shows the invariant $\gamma\gamma$ mass
distribution of all photon pair combinations for selected $9\gamma$
events.  A clear \piz signal is visible.  By combining four \piz
candidates with an additional photon candidate being detected in the
EMC barrel, where the same photon candidate must not be used more than
once, the complete event is reconstructed.
Different pairings of the photon candidates can yield more than one
$\gamma\fourpiz$ candidate per event. Therefore we use the pairing
which leads to the minimal

\begin{equation}
\chi^2_{4\pi} = \sum_i \frac{\left(m_{\gamma\gamma,i}-m_{\piz}\right)^2}{\sigma^2}
\end{equation}
calculated from the invariant mass $m_{\gamma\gamma,i}$ of the
$i^{\mathrm{th}}$ \piz candidate for a given $\gamma\fourpiz$
candidate, the nominal \piz mass \cite{pdg} $m_{\piz}$, and the
$\piz\to\gamma\gamma$ invariant mass resolution $\sigma$ of 6.5\mevcc. Combinatorial background is suppressed strongly
by demanding $\chi_{4\pi}^2<15$.

Potential backgrounds can arise from the transition
$\psiprime\to\piz\piz\jpsi$ followed by hadronic or radiative decays
of the \jpsi to final states with higher photon multiplicity. We
therefore require the recoil mass $m_{R}$ of any di-\piz
pair with respect to the \psiprime to be
$|m_{R}-m_{\jpsi}|>100\mevcc$, where $m_{\jpsi}$ is the nominal
\jpsi mass \cite{pdg}.

\begin{figure}[tb]
\includegraphics[width=1\columnwidth]{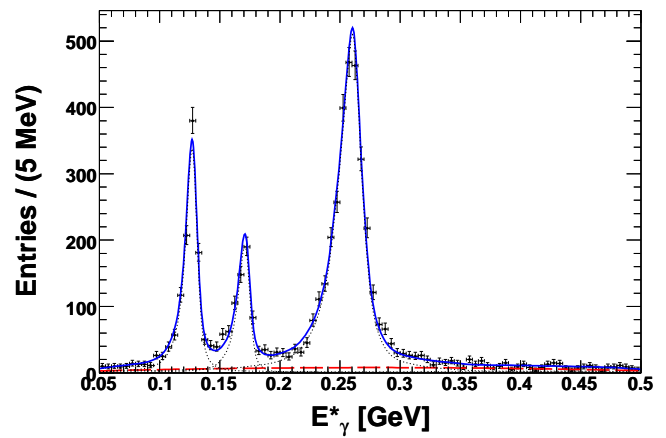}
\caption{The spectrum of the energy \Erad of the radiative photon from
  $\psiprime\to\gamma\chic{J}$ with the result of the fit (solid curve) described in the text. The dashed curve
  shows the background line shape and the dotted curves represent the
  signal line shapes derived from MC simulations (see
  Sect.~\ref{sec:fitting}).}
\label{fig:radgamma}
\end{figure}

The spectrum of the energy $E_\gamma^*$ of the photon from the
\psiprime radiative transition in the center of mass frame is shown in
Fig.~\ref{fig:radgamma}.  Clear \chic{0}, \chic{1},and \chic{2}
signals with small background are evident.  Analysis of the continuum
data sample yields only four events passing the selection criteria and
thus reveals no significant background.  Peaking components of the
background are investigated from simulated Monte Carlo (MC) events and
are discussed in Sec.~\ref{sec:mc}.

\begin{figure*}[hbt]
\begin{center}
\includegraphics[width=\textwidth]{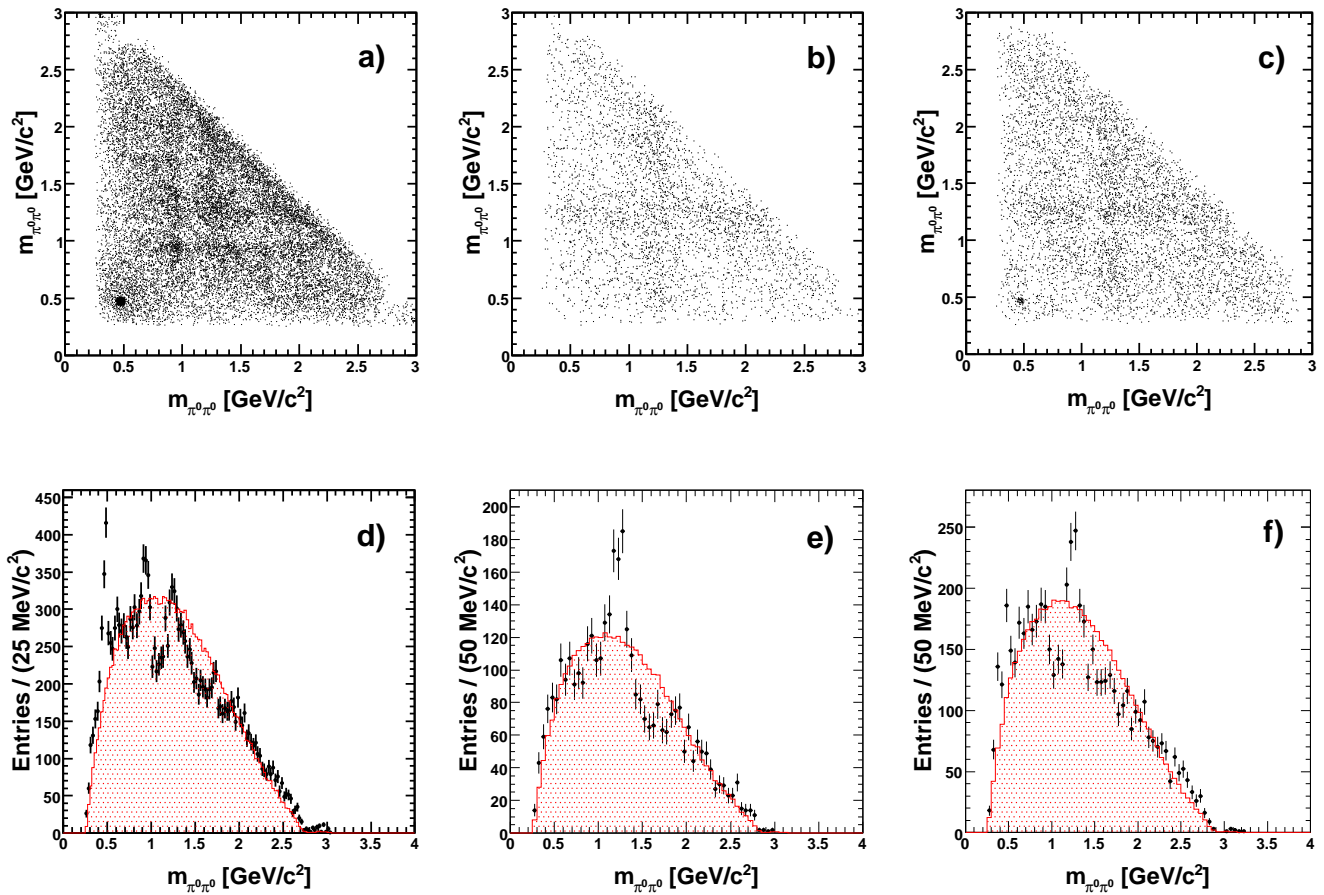}
\caption{Shown is the invariant di-\piz mass plotted versus the other
  invariant di-\piz mass for events selected from the a) \chic0, b)
  \chic1, and c) \chic{2} signal regions. All possible \piz\piz
  combinations of the \fourpiz hadronic final state are plotted. In
  addition the plots are symmetrized; thus each event enters six times
  to the plots. The one-dimensional projections are shown in d) e) and
  f) for \chic0, \chic1, and \chic{2}, respectively, where the dots
  with error bars show the data and the shaded histograms show the
  di-\piz mass distributions obtained from signal MC events simulated
  without intermediate resonances for the \chic{J} decays.}
\label{fig:substructure}
\end{center}
\end{figure*}

We further look into resonant substructures in the \fourpiz final
state. The production of intermediate resonances in the decay could
have an impact on the detection efficiencies. Here, we only
investigate the gross substructures by plotting the invariant mass of
any di-\piz pair in the final state versus the corresponding mass of
the other pair for the three \chic{J} signal regions
(Fig.~\ref{fig:substructure}). The defined \chic{0}, \chic{1}, and \chic{2}
signal regions correspond to $E_\gamma^*$ energy ranges of 220-290\mev,
160-180\mev, and 115-135\mev, respectively. As seen in
Fig.~\ref{fig:substructure} production of \KS and \fz980 in the \chic0
decay is evident. Accumulation of events is also observed in the mass
region around 1300, 1700, and 1950\mevcc. Structures around
1300\mevcc are observed in \chic{1} and \chic{2} decays. All these
structures need further careful investigation using partial wave
analysis techniques with increased data samples being collected in the
future. As for \chic{0} decays, \KS production is also observed in
\chic{2} decays, while the decay $\chic1\to\KS\KS$ is forbidden by
parity-conservation.

For the measurement of branching fractions we include all
sub-resonant decay modes but explicitly exclude the \chic{0} and
\chic{2} decay mode to $\KS\KS$. Therefore we reject events
where the invariant mass $m_{12}$ of any di-\piz pair and 
the invariant mass $m_{34}$ of the corresponding other di-\piz pair 
of the $\gamma\fourpiz$ final state fulfills 
$\sqrt{(m_{12}-m_{\KS})^2+(m_{34}-m_{\KS})^2 }<100\mevcc$, 
where $m_{\KS}$ is the nominal~\KS mass \cite{pdg}.

\section{Monte Carlo Studies}
\label{sec:mc}
A detailed MC simulation of the \besiii detector based
on {\sc geant4} \cite{geant4} is used to determine efficiencies, signal shapes, and
background contributions. The production of the \psiprime resonance is
simulated using the {\sc kkmc} event generator \cite{kkmc}. Decays of
the \psiprime and subsequent particles in the event are modeled by
{\sc evtgen} \cite{evtgen}. Simulated events pass the same reconstruction
algorithms and selection criteria as data.

Signal MC data samples of 500k events for each decay
$\psiprime\to\gamma\chic{J}$, $\chic{J}\to\fourpiz$ are generated
using a $1+\lambda\cos^2(\theta)$ distribution, where $\theta$ is the angle between the direction of the radiative photon and the positron
beam, and $\lambda=1,-1/3,1/13$ for $J=0,1,2$ in accord with
expectations for electric dipole (E1) transitions. The \chic{J} decay
products are generated using a flat angular distribution. Intrinsic
width and mass values as given in \cite{pdg} are used for the
\chic{J} states in the simulation. The obtained efficiencies for
\chic{0}, \chic{1}, and \chic{2} are $(10.16 \pm 0.05)\%$,
$(11.54 \pm 0.05) \%$, and $(10.85 \pm 0.05) \%$, respectively, including detector acceptance as well as reconstruction 
and selection 
efficiencies.

In addition we use MC data samples to investigate sources of the peaking backgrounds. 
For each of the studied \chic{J} decay modes listed in Table
\ref{tab:peakingbg} we generated at least 100k events.  The
contribution of the total peaking
background is estimated from a fit to the reconstructed \Erad
spectrum. The fit procedure is the same as applied for data and will
be addressed in Sect.~\ref{sec:fitting}. The largest peaking
background contribution is found to come from $\chic1\to\eta\piz\piz$
decays, with $\eta\to\piz\piz\piz$, where one of the \piz has low
momentum and is not detected.

\begin{table}[tb]
\begin{center}
\begin{tabular} {lccc}
\hline \hline
channel & $n_{\chic{0}}$ & $n_{\chic{1}}$ & $n_{\chic{2}}$  \\\hline
$\chic{0} \to \KS \KS, \KS\to\piz\piz$   &  1.6 & 0.3 & 0\\
$\chic{0} \to \eta \eta$  &  0.2 & 0 & 0 \\ \hline
$\chic{1} \to \eta \piz \piz$   &  1.2 & 45.2 & 0\\
$\chic{1} \to \gamma \jpsi, \jpsi\to \omega \piz \piz$    &  0 & 1.3 & 0 \\
$\chic{1} \to \gamma \jpsi, \jpsi \to \eta \piz \piz$   &  0 & 0 & 0.1  \\ \hline
$\chic{2} \to \KS \KS, \KS \to \piz\piz$   &  0 & 0 & 0.6 \\
$\chic{2} \to \eta \piz \piz$   &  0 & 0 & 3.8 \\
$\chic{2} \to \eta \eta$   &  0 & 0 & 0 \\
$\chic{2} \to \gamma \jpsi, \jpsi \to \omega \piz \piz$   &  0 & 0 & 0.6 \\
$\chic{2} \to \gamma \jpsi, \jpsi \to \eta \piz \piz$    &  0 & 0  & 0 \\
\hline
Sum & 3.0& 46.8 & 5.1 \\
\hline\hline
\end{tabular}
\caption{Expected number of background events peaking at the \chic{J} signal
  regions as derived from MC simulations.}
\label{tab:peakingbg}
\end{center}
\end{table}

\section{Fitting Procedure and Extraction of Branching Fractions}
\label{sec:fitting}
The \Erad spectrum shown in Fig.~\ref{fig:radgamma} is fitted using an
unbinned maximum likelihood fit. The \chic{J} signal line shapes are
extracted from the MC simulation. A 2$^\mathrm{nd}$ order Chebychev
polynomial is used to describe the non-peaking background. From the
fit \chic{0}, \chic{1}, and \chic{2} signal yields of $3299\pm67$,
$655\pm32$, and $1169\pm41$, respectively, are obtained.

To access the goodness of the fit, we repeat the fit to the \Erad spectrum using a binned least-squared fit. 
Applying a binning of 5\mev, this fit yields a $\chi^2$ value of 102 with 84 degrees of freedom.

The fit does not account for the peaking component of the
background. We estimate the number of events peaking at the position
of the \chic{J} signals by fitting the \Erad spectrum derived from the
MC samples generated for background studies. The same fitting
procedure as for data is applied, except that the parameters of the
polynomial function describing the shape of the non-peaking
combinatorial background are fixed to the values obtained from the fit
to data. From the extracted signal yields the expected number of
peaking background events is calculated using the branching fractions
of the $\chi_{cJ}$ decays as given in \cite{pdg}.  We relate
the unmeasured branching fractions of the decays
$\chic{J}\to\eta\piz\piz$ to the corresponding branching fractions of
$\chic{J}\to\eta\pip\pim$ decays by the iso-spin ratio of the two
final states. The estimated number of peaking background events for
the investigated channels is given in Table~\ref{tab:peakingbg}. In
total a peaking background of 3.0, 46.8 and 5.1 events to the
\chic{0}, \chic{1}, and \chic{2} signals, respectively, is derived. The
largest contribution arises from the decay $\chic{1}\to\eta\piz\piz$,
amounting to 45 events at the \chic{1} signal region. 

Although $\chic0\to\KS\KS$ and $\chic2\to\KS\KS$ decays for the
branching fraction measurement were excluded, we considered these
channels as a potential peaking background source. The feed-through of
the two decays to the \chic{0}, \chic{1}, and \chic{2} signals is
expected to be 1.6, 0.3, and 0.6 events, respectively.

The expected number of peaking background events is subtracted from
the yields observed for data. These corrected yields $N$ are then
converted to branching fractions using
\begin{eqnarray}
 \mathcal{B}(\chic{J}\to 4 \piz) = 
\frac{N}{\epsilon\cdot N_{\psiprime}\cdot \mathcal{B}(\psiprime\to\gamma\chic{J})\cdot \mathcal{B}(\piz\to\gamma\gamma)^4  }
\end{eqnarray}
where $\epsilon$ is the efficiency; $N_\psiprime$ is the number of
\psiprime in the data sample; and
$\mathcal{B}(\psiprime\to\gamma\chic{J})$ and
$\mathcal{B}(\piz\to\gamma\gamma)$ are the branching fractions of
radiative \psiprime transitions into \chic{J} and of the decay
$\piz\to\gamma\gamma$~\cite{pdg}, respectively.  The number of
\psiprime and its combined statistical and systematical uncertainties
are determined to be $N_{\psiprime}=(1.06\pm0.04)\times10^8$
\cite{psi2scounting}.

\section {Estimation of Systematic Uncertainties}
\label{sec:systematics}
Several sources of systematic uncertainties are considered for the
measurement of the branching fractions, including uncertainties on the
photon detection and reconstruction; the event selection; the fitting
procedure and peaking background subtraction; and the number of
$\psiprime$ decays in the data sample. The investigated uncertainties
are summarized in Table~\ref{tab:systematics} and will be discussed in
detail in the following.

\begin{table}[tb]
\begin{center}
\begin{tabular} {lrrr}
\hline\hline
                               &  $\chi_{c0}[\%]$ &  $\chi_{c1}[\%]$ &  $\chi_{c2}[\%]$ \\
\hline
photon detection                            & 9.0  & 9.0  & 9.0   \\
decay model                                 & 6.3  & 6.3  & 6.3   \\
$\mathcal{B}(\psiprime\to\gamma\chic{J})$   & 3.2   & 4.3 & 4.0 \\ 
number of $\psiprime$ events                 & 4.0  & 4.0  & 4.0   \\
total energy                                & 3.0  & 3.0  & 3.0   \\   
$\chi^2_{4\pi}$                             & 2.5  & 2.5  & 2.5   \\
reconstructed $\piz$ mass                  & 1.4  & 1.4  & 1.4   \\
fitting range                               & 1.4  & 2.9  & 0.9   \\
signal line shape (energy resolution)        & 0.5  & 1.2  & 0.4   \\
signal line shape (energy shift)             & 0.1  & 0.6  & 1.0   \\
background shape                            & 1.0  & 1.0  & 1.0   \\
peaking background subtraction              & $0.1$ & $0.8$ & $0.5$ \\
MC statistics                               & 0.5  & 0.5  & 0.5   \\
trigger  efficiency                         & $<$0.1  & $<$0.1  & $<$0.1   \\
\hline
total uncertainty                           & 13.2  & 13.6  & 13.2   \\
\hline \hline
\end{tabular}
\caption{Summary of the systematic uncertainties. }
\label{tab:systematics}
\end{center}
\end{table}

\begin{paragraph}{Photon Detection}  
The uncertainty due to photon detection and conversion is 1\% per
photon. This is determined from studies of photon detection in well
understood decays such as $\jpsi\to\rho^0\piz$ and the study of photon
conversion in the process $\epem\to\gamma\gamma$.
\end{paragraph}

\begin{paragraph}{Event Selection}
By varying the requirement on $\chi_{4\pi}^2$, the \piz mass window
and the total energy of the $\gamma\fourpiz$ candidates used for the
event selection in data and MC events, we investigate the systematic
uncertainties in modeling the distribution of these parameters. The
largest deviation of the branching fractions from the default values
sets the scale of our systematic uncertainty, and we assign a
uncertainty of 2.5\% for the $\chi_{4\pi}^2$; 3\% for the total
energy; and 1.4\% for the \piz mass window requirement.
\end{paragraph}

\begin{paragraph}{Monte Carlo Decay Model}
The efficiencies for the processes $\psiprime\to\gamma\chic{J}$,
$\chic{J}\to\piz\piz\piz\piz$ are determined from MC simulations,
where no intermediate resonances and a flat angular distribution have
been considered in the $\chic{J}$ decays. As discussed in
Sect.~\ref{sec:selection} this analysis reveals the presence of
intermediate resonances in the $\chic{J}\to\fourpiz$ decays. This
could have an impact on the detection efficiencies, which we consider
as a systematic uncertainty. We determine the efficiencies from our
simulations including the sub-resonant modes $f_0(980)f_0(980)$,
$f_2(1270)f_2(1270)$, $f_0(1370)f_0(1500)$, and $f_0(1370)f_0(1710)$
for \chic0. We considered $f_2(1270)\piz\piz$ for \chic{1} and
$f_2(1270)f_2(1270)$ and $f_0(1370)f_0(1710)$ for \chic{2}. We find
there is no large efficiency difference from that of phase space.  The
largest difference with respect to the efficiency obtained for the
simulation without intermediate resonances is observed for
$\chic{0}\to f_0(1370)f_0(1710)$ decay. We take this difference as a
conservative estimate of the uncertainty due to the MC decay model and
assign a uncertainty of 6.3\% for the \chic{0}, \chic{1}, and \chic{2}
branching fractions.
\end{paragraph}

\begin{paragraph}{Fitting Procedure}
The $\chic{J}$ yields determined from the fit to the \Erad
spectrum determines the branching fractions. We repeat the fit with
appropriate modifications to estimate the systematic uncertainties due
to the fitting procedure. The difference of the derived branching
fractions with respect to the values derived from the standard fit is
considered as a systematic uncertainty.

We smear the resolution function of the $\chic{J}$ signals obtained
from MC simulations by $1\%$ to estimate the systematic uncertainties
of modeling the photon resolution in the MC simulation and shift the
signal mean values by $\pm1\mev$ to estimate the systematic
uncertainties due to the absolute energy calibration. The assigned
uncertainties are given in Table~\ref{tab:systematics}. To estimate
the uncertainty due to the non-peaking background parametrization we
use a third order instead of a second order Chebychev
polynomial. This uncertainty is found to be $1\%$. For the nominal
fit, the \Erad spectrum is fitted in the interval 0.05-0.5\gev. A
series of fits using different \Erad intervals is performed and the
largest change of the individual branching fractions is assigned as a
systematic uncertainty (see Table~\ref{tab:systematics}).
\end{paragraph}

\begin{paragraph}{Peaking Background Subtraction}
The number of peaking background events is estimated from MC simulations
and is subtracted from the signal yields obtained from the nominal
fit. The major source of peaking background is the decay
$\chic{1}\to\eta\piz\piz$. To determine the number of background
events, the branching fraction of this decay mode is computed from the
$\chic{1}\to\eta\pip\pim$ branching fraction exploiting the isospin
relation of the two decays.  The uncertainty of the branching fraction
of $\chic{1}\to\eta\piz\piz$ leads to a systematic uncertainty of 0.8\%
for \chic{1}.  For \chic{0} and \chic{2} we take the number of
subtracted background events as systematic uncertainties.
\end{paragraph}

\begin{paragraph}{Other Systematic Uncertainties}
For the normalization of the branching fractions, the number of
$\psiprime$ events in the data sample determined according to the
method as given in \cite{psi2scounting} is used. This method
yields a systematic uncertainty of $4\%$. The uncertainty due to the
branching fractions $\psiprime\to\gamma\chic{J}$ is 3.2\% for
\chic{0}, 4.3\% for \chic{1}, and 4.0\% for \chic{2}.

A systematic uncertainty of 0.5\% is assigned due to the statistical
error of the efficiencies as determined from MC simulations. 

The systematic uncertainty due to the simulation of the trigger
efficiency is found to be less than $0.1\%$ \cite{trigger}.
\end{paragraph}

\begin{paragraph}{Total Systematic Uncertainty}
We assume that all systematic uncertainties given above are
independent and add them in quadrature to obtain the total systematic
uncertainty.
\end{paragraph}

\section{Conclusion}
\label{sec:summary}
In summary we have measured the branching fractions of
$\chic{J}\to\fourpiz$ decays $\mathcal{B}(\chic{0}\to\fourpiz)=
(3.34\pm 0.06 \pm 0.44)\times10^{-3}$,
$\mathcal{B}(\chic{1}\to\fourpiz)= (0.57\pm 0.03 \pm
0.08)\times10^{-3}$, $\mathcal{B}(\chic{2}\to\fourpiz) = (1.21\pm 0.05
\pm 0.16)\times10^{-3}$ for the first time, where the quoted
uncertainties are statistical and systematical, respectively. The
\fourpiz hadronic final state contains a rich substructure of
intermediate resonances. The reported branching fractions include
decay modes with intermediate resonances except $\chic{0}\to\KS\KS$
and $\chic{2}\to\KS\KS$, which have been removed from this
measurement. Our observation improves the existing knowledge of the
\chic{J} states and provides further insight into their decay
mechanisms. Based on our results detailed studies of the sub-resonant
decay structure with increased data samples may follow in the future.

\section{Acknowledgments}
The BESIII collaboration thanks the staff of BEPCII and the computing center for their hard efforts. This work is supported in part by the Ministry of Science and
Technology of China under Contract No. 2009CB825200; National Natural Science Foundation of China (NSFC) under Contracts Nos. 10625524, 10821063, 10825524, 10835001,
10935007; the Chinese Academy of Sciences (CAS) Large-Scale Scientific Facility Program; CAS under Contracts Nos. KJCX2-YW-N29, KJCX2-YW-N45; 100 Talents Program of
CAS; Istituto Nazionale di Fisica Nucleare, Italy; Russian Foundation for Basic Research under Contracts Nos. 08-02-92221, 08-02-92200-NSFC-a; Siberian Branch of
Russian Academy of Science, joint project No 32 with CAS; U. S. Department of Energy under Contracts Nos. DE-FG02-04ER41291, DE-FG02-91ER40682, DE-FG02-94ER40823;
University of Groningen (RuG) and the Helmholtzzentrum fuer Schwerionenforschung GmbH (GSI), Darmstadt; WCU Program of National Research Foundation of Korea under
Contract No. R32-2008-000-10155-0

\end{document}